\documentclass[conference, final, letterpaper]{IEEEtran}

\usepackage[utf8]{inputenc}
\usepackage{graphicx}
\usepackage{stfloats}
\usepackage{amssymb}
\usepackage[cmex10]{amsmath}
\usepackage{amsthm} 
\usepackage{array}
\usepackage[usenames,dvipsnames]{xcolor}
\usepackage{hyperref}
\hypersetup{colorlinks=false, pdfborderstyle={/S/U/W 1},pdfborder=0 0 1, citebordercolor=Blue, filebordercolor=Red, linkbordercolor=Red, urlbordercolor=Blue}
\usepackage{cite}
\usepackage{xparse}
\usepackage{nicefrac}
\usepackage[english]{babel}
\usepackage[english = american]{csquotes}
\usepackage{tabularx}
\newtheorem{definition}{Definition}
\newtheorem{theorem}[definition]{Theorem}
\newtheorem{lemma}[definition]{Lemma}

\newtheorem{example}[definition]{Example}

\newtheorem{construction}[definition]{Construction}

\DeclareMathOperator{\defi}{def}
\newcommand{\defeq}{\overset{\defi}{=}}

\newcommand{\F}[1]{\mathbb F_{#1}}
\newcommand{\Fq}{\F{q}}
\newcommand{\Fxsub}[1]{\ensuremath{\mathbb{F}_{#1}[X]}}
\newcommand{\Fqx}{\Fxsub{q}}

\newcommand{\Z}{\ensuremath{\mathbb{Z}}}

\newcommand{\M}[2][\empty]{
  \ifthenelse{\equal{#1}{\empty}}
    {\ensuremath{\mathbf{#2}}}
    {\ensuremath{{\mathbf{#2}}_{#1}}}
}

\newcommand{\CYC}{\ensuremath{\mathcal{C}}}
\newcommand{\ADD}{\ensuremath{\mathcal{A}}}
\newcommand{\LOC}{\ensuremath{\mathcal{L}}}
\newcommand{\LIN}[4]{\ensuremath{[#1,#2,#3]_{#4}}}

\newcommand{\interval}[1]{\ensuremath{[#1)}}
\newcommand{\CYCn}{\ensuremath{n}}
\newcommand{\CYCk}{\ensuremath{k}}
\newcommand{\CYCd}{\ensuremath{d}}
\newcommand{\CYCext}{\ensuremath{m}}
\newcommand{\ADDn}{\ensuremath{n'}}
\newcommand{\ADDk}{\ensuremath{k'}}
\newcommand{\ADDd}{\ensuremath{d'}}
\newcommand{\ADDext}{\ensuremath{m'}}
\newcommand{\LOCn}{\ensuremath{n_l}}
\newcommand{\LOCk}{\ensuremath{k_l}}
\newcommand{\LOCd}{\ensuremath{d_l}}

\newcommand{\coset}[2]{\ensuremath{M_{#1,#2}}}
\newcommand{\defset}[2][\empty]{
  \ifthenelse{\equal{#1}{\empty}}
    {\ensuremath{D_{#2}}}
    {\ensuremath{D^{[#1]}_{#2}}}
}
\newcommand{\NONLIN}[4]{\ensuremath{(#1,#2,#3)_{#4}}}
\addto\captionsenglish{}
\begin{document}
\title{Optimal Linear and Cyclic Locally Repairable Codes over Small Fields}

\IEEEoverridecommandlockouts

\author{\IEEEauthorblockN{Alexander Zeh and Eitan Yaakobi}\thanks{This work has been supported by German Research Council (Deutsche Forschungsgemeinschaft, DFG) under grant ZE1016/1-1.}
\IEEEauthorblockA{Computer Science Department\\
Technion---Israel Institute of Technology\\
\texttt{alex@codingtheory.eu}, \texttt{yaakobi@cs.technion.ac.il}}
}

\maketitle

\begin{abstract}
We consider locally repairable codes over small fields and propose constructions of optimal cyclic and linear codes in terms of the dimension for a given distance and length.

Four new constructions of optimal linear codes over small fields with locality properties are developed.
The first two approaches give binary cyclic codes with locality two. While the first construction has availability one, the second binary code is characterized by multiple available repair sets based on a binary Simplex code. 

The third approach extends the first one to $q$-ary cyclic codes including (binary) extension fields, where the locality property is determined by the properties of a shortened first-order Reed--Muller code. 
Non-cyclic optimal binary linear codes with locality greater than two are obtained by the fourth construction.
\end{abstract}

\section{Introduction}
Locally repairable codes (LRC) can recover from erasure(s) by accessing a small number of erasure-free code symbols and therefore increase the efficiency of the repair-process in large-scale distributed storage systems.
Basic properties and bounds of LRCs were identified by Gopalan~\textit{et al.}~\cite{gopalan_locality_2012}, Oggier and Datta~\cite{oggier_self-repairing_2011} and Papailiopoulos and Dimakis~\cite{papailiopoulos_locally_2014}. 
The majority of the constructions of LRC requires a large field size (see e.g.~\cite{tamo_optimal_2013, silberstein_optimal_2013, tamo_family_2014}). 
The work of Kuijper and Napp~\cite{kuijper_erasure_2014} considers binary LRCs (and over binary extension field). Cadambe and Mazumdar~\cite{cadambe_upper_2013-1} gave an upper bound on the dimension of a (nonlinear) code with locality which takes the field size into account.
Goparaju and Calderbank~\cite{goparaju_binary_2014} proposed binary cyclic LRCs with optimal dimension (among linear codes) for distances $6$ and $10$ and locality $2$.

Our paper is based on the work of Goparaju and Calderbank~\cite{goparaju_binary_2014} and we use their projection to an additive code without locality (see Calderbank~\textit{et al.}~\cite{calderbank_quantum_1998}, Gaborit~\textit{et al.}~\cite{gaborit_additive_2001}, Kim~\textit{et al.}~\cite{kim_projections_2003} for additive codes). We construct a new family of optimal binary codes (with distance $10$ and locality $2$) and generalize the approach to $q$-ary alphabets. 
Furthermore, we give a construction of optimal binary cyclic codes with availability greater than one based on Simplex codes (see Pamies-Juarez~\textit{et al.}~\cite{pamies-juarez_locally_2013}, Rawat~\textit{et al.}~\cite{rawat_locality_2014} for the definition of availability and Kuijper and Napp~\cite{kuijper_erasure_2014} for a Simplex code based construction).

This paper is structured as follows. Section~\ref{sec_Preliminaries} gives necessary preliminaries on linear and cyclic codes, defines LRCs, recalls the generalized Singleton bound, the Cadambe--Mazumdar bound~\cite{cadambe_upper_2013-1} as well as the definition of availability for LRCs.
The concept of a locality code and the projection to an additive code are discussed in Section~\ref{sec_LocalityCode} based on the work of Goparaju and Calderbank~\cite{goparaju_binary_2014}. Two new constructions of optimal binary codes are given in Section~\ref{sec_BinaryCyclicCodes} and a construction based on a $q$-ary shortened cyclic first-order Reed--Muller code is given in Section~\ref{sec_ReedMuller}. The fourth construction in Section~\ref{sec_LinearCodes} uses code concatenation and provides optimal linear binary codes. Section~\ref{sec_conclusion} concludes this contribution.

\section{Preliminaries} \label{sec_Preliminaries}
Let $\interval{a,b}$ denote the set of integers $\{a,a+1,\dots,b-1\}$ and $\interval{b}$ be the shorthand notation for $\interval{0,b}$.
Let $\Fq$ denote the finite field of order $q$ and $\Fqx$ the polynomial ring over $\Fq$ with indeterminate $X$. 
A linear $\LIN{n}{k}{d}{q}$ code of length $n$, dimension $k$ and minimum Hamming distance $d$ over $\Fq$ is denoted by a calligraphic letter like $\CYC$ as well as a non-linear $\NONLIN{n}{M}{d}{q}$ of length $n$, cardinality $M$ and minimum distance $d$.

An \LIN{n}{k}{d}{q} $q$-ary cyclic code $\CYC$ with distance $d$ is an ideal in the ring $\Fqx / (X^n-1)$ generated by $g(X)$. 
The generator polynomial $g(X)$ has roots in the splitting field $\F{q^s}$, where $n
\mid (q^s -1)$.

A $q$-cyclotomic coset $\coset{i}{n}$ is defined as 
\begin{equation} \label{eq_cyclotomiccoset}
\coset{i}{n} \defeq \big\{ iq^j \mod n \, \vert \, j \in \interval{a} \big\},
\end{equation}
where $a$ is the smallest positive integer such that $iq^{a} \equiv i \bmod n$. 
The minimal polynomial in $\Fqx$ of the element $\alpha^i \in \F{q^{\CYCext}}$ is given by $m_i(X) = \prod_{j \in \coset{i}{n}} (X-\alpha^j)$.
The defining set $\defset{\CYC} $ of an \LIN{n}{k}{d}{q} cyclic code
$\CYC$ is
\begin{equation} \label{eq_definingset}
\defset{\CYC}  =  \big\{ 0 \leq i \leq n-1 \, | \, g(\alpha^i)=0 \big\}.
\end{equation}
For visibility we sometimes mark a position $i$ with a $\square$ if $g(\alpha^i) \neq 0$.
Furthermore, let $\defset[z]{\CYC}$ be the short-hand notation for $\{ (i+z) \; | \; i \in \defset{\CYC} \}$ for a given $z \in \Z$.
Let us recall the definition of linear locally repairable codes.
\begin{definition}[Locally Repairable Code (LRC)] \label{def_Locality}
A linear \LIN{n}{k}{d}{q} code $\CYC$ is said to have $(r, \delta)$-locality if for all $n$ code symbols $c_i, \forall i \in \interval{n}$, there exists a punctured subcode of $\CYC$ with support containing $i$, whose length is at most $r + \delta - 1$, and whose minimum distance is at least $\delta$.
\end{definition}
A code $\CYC$ is called \emph{$r$-local} if it has $(r,2)$-locality.

The following generalization of the Singleton bound for LRCs was among others proven in~\cite[Thm. 3.1]{kamath_codes_2014}, \cite[Construction 8 and Thm. 5.4]{tamo_family_2014} and \cite[Thm. 2]{prakash_optimal_2012}.
\begin{theorem}[Generalized Singleton Bound] \label{theo_GenSingleton}
The minimum distance $d$ of an \LIN{n}{k}{d}{q} linear $(r, \delta)$-locally repairable code $\CYC$ (as in Def.~\ref{def_Locality}) is upper bounded by
\begin{equation} \label{eq_GeneralizedSingleton}
d \leq n-k+1-\left( \left \lceil \frac{k}{r} \right \rceil-1 \right)(\delta -1).
\end{equation}
\end{theorem}
For $\delta=2$ and $r=k$ it coincides with the classical Singleton bound.
Throughout this contribution we call a code \textit{Singleton-optimal} if its distance meets the bound in Thm.~\ref{theo_GenSingleton} with equality.
The generalized Singleton bound as in Thm.~\ref{theo_GenSingleton} does not take the field size into account. We compare our constructions with the bound given by Cadambe and Mazumdar~\cite[Thm. 1]{cadambe_upper_2013-1} which depends on the alphabet size. In general it holds also for nonlinear codes, but we state it only for linear codes in the following.
\begin{theorem}[Cadambe--Mazumdar (CM Bound)] \label{theo_CMBound}
The dimension $k$ of an $r$-local repairable code $\CYC$ of length $n$ and minimum Hamming distance $d$ is upper bounded by
\begin{equation} \label{eq_CMBound}
k \leq \min_{t \in \Z} \left\{tr + k_{opt}^{(q)}\big( n-t(r+1), d \big)  \right\},
\end{equation}
where $k_{opt}^{(q)}(n, d)$ is the largest possible dimension of a code of length $n$, for a given alphabet size $q$ and a given minimum distance $d$.
\end{theorem}
In the following we use Thm.~\ref{theo_CMBound} to bound the dimension of linear codes.
Another important parameter for LRCs is the availability e.g. considered in Kuijper and Napp~\cite{kuijper_erasure_2014} and Cadambe and Mazumdar~\cite{cadambe_upper_2013-1} and therefore we define it in the following.
\begin{definition}[Availability] \label{def_Availability}
An \LIN{n}{k}{d}{q} linear code $\CYC$ is called $t$-available-$r$-local locally repairable if every code symbol $c_i, \forall i \in \interval{n}$, has at least $t$ parity-checks of weight $r+1$ which intersect pairwise in (and only in) $\{i\}$.
\end{definition}

\section{Locality Code and Additive Code} \label{sec_LocalityCode}
In this section we shortly recall the approach of Goparaju and Calderbank~\cite{goparaju_binary_2014} and extend it to what we call a \textit{locality code}.

Let us first describe the idea of Constructions 1 and 2 of \cite{goparaju_binary_2014} in terms of a locality code.
Construction~1 of \cite{goparaju_binary_2014} gives a binary cyclic code $\CYC$ of length $n = 2^m-1$ with locality $r$, where the code length $n$ is divisible by $r+1$. The defining set is $\defset{\CYC} = \{ i \mod (r+1), \forall i \in \interval{n} \}$. This equals the union of $n/(r+1)$ shifted defining sets $\defset{\LOC} = \{0\}$ of the binary cyclic $\LIN{r+1}{r}{2}{2}$ single-parity check code, which is able to correct one erasure within a block of length $r+1$ by ``accessing'' only $r$ other code symbols. The code $\CYC$ inherits the properties of $\LOC$, also the minimum distance of two, which is Singleton-optimal, but does not increase the overall erasure-correction capability. 
In general, let $\CYC$ be the aimed $\LIN{n}{k}{d}{q}$ code with locality properties that are inherited from an $\LIN{\LOCn}{\LOCk}{\LOCd}{q}$ locality code $\LOC$.
Namely the code constructions of \cite{goparaju_binary_2014} and our (cyclic) codes are subcodes of the $\LIN{n}{k}{d}{q}$ cyclic product code $\LOC \otimes \mathcal{T}$, where $\mathcal{T}$ is the trivial $\LIN{n/\LOCn}{n/\LOCn}{1}{q}$ code (see~\cite{burton_cyclic_1965}), i.e., a cyclic code with defining set:
\begin{equation} \label{eq_DefsetCyclic}
\defset{\CYC} = \left\{ \defset{\LOC} \cup \defset[\LOCn]{\LOC} \cup \dots \cup \defset[n-\LOCn-1]{\LOC} \cup R \right\}.
\end{equation}
In Construction 2 ($R = \coset{1}{n}$) and 3 ($R = \coset{1}{n} \cup \coset{-1}{n} $) of \cite{goparaju_binary_2014}, the locality code $\LOC$ is a $\LIN{3}{2}{2}{2}$ single-parity check code with defining set $\defset{\LOC} = \{0\}$. The optimality among binary codes with locality $r=2$ is shown via the projection to an additive code.
\begin{lemma}[Projection to Additive Code] \label{lem_ProjectionAdditiveCode}
Let $\LOC$ be an $\LIN{\LOCn}{\LOCk}{\LOCd}{q}$ locality code and let $\CYC$ be an $\LIN{n}{k}{d}{q}$ code with defining set as in~\eqref{eq_DefsetCyclic}.
Then, we can project each sub-block of $\LOCn$ symbols of a codeword in $\CYC$ to one symbol in $\F{q^{\LOCk}}$. The obtained $\NONLIN{\ADDn}{q^k}{\ADDd}{q^{\LOCk}}$ additive code $\ADD$ has parameters:
\begin{equation} \label{eq_GeneralMapping}
\ADDn = \CYCn/\LOCn \quad \text{and} \quad \ADDd \geq \lceil d/\omega \rceil, 
\end{equation}
where $\omega$ is the maximum weight of a codeword in $\LOC$.
\end{lemma}
\begin{IEEEproof}
The length $\ADDn$ and the alphabet-size follow directly from the projection of the coordinates. The cardinality of $\ADD$ equals the one of $\CYC$. The distance follows from the fact that in the worst-case $\omega$ non-zero symbols of $\CYC$ are projected to one symbol over $\F{q^{\LOCk}}$ (see Fig.~\ref{fig_LocalityCode}).
\end{IEEEproof}
\begin{figure}[htb]
\begin{center}
\includegraphics[width=.8\columnwidth]{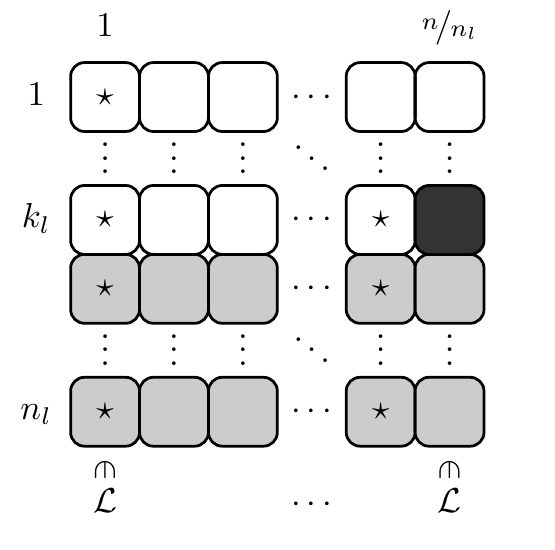}
\end{center}
\label{fig_LocalityCode}
\caption{Illustration of a nonzero minimum-weight codeword of weight seven of the $\LIN{n}{k}{d}{q}$ code $\CYC$ arranged in $\CYCn/\LOCn$ blocks of length $\LOCn$. The $\star$ marks a nonzero symbol in $\Fq$. The corresponding codeword of the additive code $\ADD$ over $\F{q^{\LOCk}}$ has length $\CYCn/\LOCn$ and has at least weight $\lceil \CYCd/\omega \rceil$. Here $\omega = 4$ and therefore at least two symbols in $\ADD$ are nonzero (first and before last column).
The redundancy added by the $\LIN{\LOCn}{\LOCk}{\LOCd}{q}$ locality code $\LOC$ is illustrated as gray symbols, while the black symbol marks the additional redundancy to obtain a distance that is higher than the one given by the Singleton bound.}
\label{fig_LocalityCode}
\end{figure}
For a cyclic binary $\LIN{r+1}{r}{2}{2}$ single-parity check code of odd length, the maximum weight of a codeword is $\omega = r$. 
\begin{lemma}[Locality Code] \label{lem_MDSLocalityCode}
If $\LOC$ is an $\LIN{\LOCn = r+\delta-1}{r}{\delta}{q}$ MDS locality code, then the cyclic code $\CYC$ with defining set as in~\eqref{eq_DefsetCyclic} has $(r,\delta)$-locality and its distance is $d \geq \delta$.
\end{lemma}
\begin{IEEEproof}
The $(r,\delta)$-locality of $\CYC$ follows directly from the construction. The distance of $\CYC \subseteq \LOC \otimes \mathcal{T}$ is at least the distance of the product code $\LOC \otimes \mathcal{T}$.
\end{IEEEproof}
Note that for $R = \emptyset$, the code $\CYC = \LOC \otimes \mathcal{T}$ is Singleton-optimal, i.e., the dimension of $\CYC$ is $k = nr/(r+\delta-1)$ and from \eqref{eq_GeneralizedSingleton} we obtain:
\begin{align*}
d & \leq n - \frac{nr}{r+\delta-1} + 1 - \left( \frac{n}{r+\delta-1} -1 \right) (\delta-1) \\
 & \leq \frac{n(r+\delta-1) - nr - n(\delta-1)}{r+\delta-1} + \delta = \delta.
\end{align*}

\section{Binary Cyclic Codes With Locality Two} \label{sec_BinaryCyclicCodes}
Construction 1 in \cite{goparaju_binary_2014} gives Singleton-optimal binary cyclic codes (these codes are of lowest-rate for $d=2$ see \cite[Prop. 2]{zeh_new_2014}).
The following construction gives a new class of binary cyclic $2$-local codes. 
\begin{construction}[Binary Reversible Codes] \label{constr_reversiblecodes}
Let $n = 2^m+1$ and $3 | n$ and therefore $m$ odd. Let the locality $r=2$, i.e. let $\LOC$ be a $\LIN{3}{2}{2}{2}$ single-parity check code with $\defset{\LOC} = \{0\}$.
Let the defining set be 
\begin{align*}
\defset{\CYC} & = \big\{ \{\dots,-6,-3,0,3,6,9,12,\dots\} \cup \coset{1}{n} \big\}, \\
& = \{ \dots, -6, -4,-3,-2,-1,0,1,2,3,4,6,\dots \}.
\end{align*}
Then $\CYC$ has dimension $k = \frac{2}{3}(2^m+1) - 2m$ and distance $d \geq 10$.
\end{construction}
Due to the length, the coset $\coset{1}{n}$ is reversible (see \cite{massey_reversible_1964} for reversible codes), i.e., $\coset{1}{n} = \{1,2,\dots,2^{m-1},2^m = -1,-2,\dots,-2^{m-1} \}$ and has cardinality $2m$.
The distance follows from the BCH bound~\cite{hocquenghem_codes_1959, bose_class_1960}, where the consecutive sequence is $-4,-3,\dots, 3,4$.
\begin{theorem} \label{theo_ConstrReversibleCodes}
Let a binary linear code with parameters as in Construction~\ref{constr_reversiblecodes} be given.
Then its dimension satisfies:
\begin{equation}
k \leq \frac{2}{3}(2^m+1) - 2m.
\end{equation}
\end{theorem}
\begin{IEEEproof}
Equivalent to the proof of \cite[Thm. 2]{goparaju_binary_2014}, we have via sphere-packing bound (see \cite[Ch. 1 \S 5]{macwilliams_theory_1988}) for the $\NONLIN{2n/3}{2^{k}}{5}{2^2}$ additive code that 
\begin{equation}
2^k \leq \frac{4^{\ADDn}}{1+3 \ADDn + \frac{9 \ADDn(\ADDn-1)}{2}}
\end{equation}
and therefore
\begin{align}
k & \leq \log_2(4^{n/3}) - \log_2 \left( 1 - \frac{1}{2}n + \frac{1}{2}n^2 \right) \nonumber  \\
 & = \frac{2n}{3} +1 - \lceil \log_2 \left(2 - n + n^2 \right) \rceil. \label{eq_BeforeInsertingLength} 
\end{align}
With $n=2^m+1$, we obtain from~\eqref{eq_BeforeInsertingLength}
\begin{align*}
k & = \frac{2n}{3} +1 - \lceil \log_2 \left(2 - (2^m+1) + (2^m+1)^2 \right) \rceil \\
 & = \frac{2n}{3} +1 - \lceil \log_2 \left(2 + 2^m + 2^{2m} \right) \rceil \\
 & = \frac{2n}{3} +1 - (2m + 1) = \frac{2n}{3} - 2m.
\end{align*}
\end{IEEEproof}
\textbf{Remark 1:} A binary cyclic code as in Construction~\ref{constr_reversiblecodes} without $\coset{1}{n}$ in the defining set is Singleton-optimal and the distance equals $d=2$ (for $k=2n/3$, $r=2$ and $\delta =2$), which is the smallest minimum distance possible for a binary cyclic code with rate $2/3$ (see \cite[Prop. 2]{zeh_new_2014}).\\
\textbf{Remark 2:} The $n$th root of unity is in $\F{2^{2m}-1}$ (which is twice the extension order of the code obtained via \cite[Construction 3]{goparaju_binary_2014}) and therefore the (non-local) decoding complexity is higher than \cite[Construction 3]{goparaju_binary_2014}.
\begin{example}[Optimal $2$-Local Binary Code]
Let $n = 2^5+1 = 33$ and via Construction~\ref{constr_reversiblecodes} we obtain a binary cyclic code of minimum distance $d=10$, with
\begin{equation*}
\defset{\CYC} = \big\{ \{ 0,3,6,9,\dots,30\} \cup \{1, 2, 4, \dots, 32\} \big\} 
\end{equation*}
and dimension $k = 12$, which is $2$-local. 
The CM bound (see Thm.~\ref{theo_CMBound}) based on the best-known linear codes give $k \leq 13$.
\end{example}
Let us consider Construction 4 of~\cite{goparaju_binary_2014} based on the $\LIN{7}{3}{4}{2}$ locality code $\LOC$ with locality $r=2$, availability $t = 3$ and with defining set $\defset{\LOC} = \big\{ 0,\square, \square,3,\square,5,6 \big\}$.
The code $\CYC$ with defining set as in~\eqref{eq_DefsetCyclic}, but with $R = \emptyset$ is an $\LIN{n=2^m-1}{k = 3n/7}{4}{2}$ cyclic code, where $m$ is a multiple of three.

We extend Construction 4 of \cite{goparaju_binary_2014} to obtain a higher distance and small reduction of the rate as follows.
\begin{construction}[Sphere-Packing Optimal Binary Code with Locality Two and Increased Availability] \label{const_gclast}
Let $n = 2^m-1$ and be divisible by 7 and therefore $3|m$.
Let the defining set be:
\begin{align*}
\defset{\CYC} & = \big\{ \{\dots,-9,-8,|-7,\square, \square,-4,\square,-2,-1,|0,\square, \square,\\ 
 & \qquad \quad 3,\square,5,6,|7,\square, \square,10,\square, \square, 12,13|,\dots\} \\
 & \qquad \quad  \cup  \coset{1}{n} = \{1,2,4,\dots,2^{m-1} \} \big\}.
\end{align*}
Then $d \geq 12$ (via BCH bound~\cite{hocquenghem_codes_1959, bose_class_1960}, where the consecutive sequence is $-2,1,\dots,8$).
The constructed $\LIN{n=2^m-1}{k = 3n/7-m}{d \geq 12}{2}$ cyclic code $\CYC$ is a $2$-local code and has availability $t=3$ as defined in Def.~\ref{def_Availability}.
\end{construction}
\begin{example}[$2$-Local Binary Code with Availability Three]
Let $n = 9 \cdot 7= 63$, $\defset{\LOC} = \big\{ 0,\square, \square,3,\square,5,6 \big\}$  and let the defining set be 
\begin{align*}
\defset{\CYC} & = \{ \defset{\LOC} \cup \defset[7]{\LOC} \cup \dots \cup \defset[56]{\LOC} \cup \coset{1}{63} \} \\
& = \{ \{..,59,61,62,|0,3,5,6,|7,10,..\} \cup \{1,2,4,..,32\} \}.
\end{align*}
The constructed code $\CYC$ is an $\LIN{63}{21}{12}{2}$ code and the corresponding additive code according to Lemma~\ref{lem_ProjectionAdditiveCode} is a $\NONLIN{9}{2^7}{3}{2^3}$ code.
The BCH bound is tight and the consecutive sequence ranges is $61,62,0,1,\dots,8$.
\end{example}
Let us prove the optimality of Construction~\ref{const_gclast} in the following theorem.
\begin{theorem}
Let $3|m$ and let $\CYC$ be an $\LIN{2^m-1}{k}{12}{2}$ linear code, let $\LOC$ be the $\LIN{7}{3}{4}{2}$ Simplex code and let $\CYC$ have the locality inherited from $\LOC$ as in Lemma~\ref{lem_ProjectionAdditiveCode}.
Then,
\begin{equation}
k \leq \frac{3}{7} \left( 2^m-1 \right)-m.
\end{equation}
\end{theorem}
\begin{IEEEproof}
From~\eqref{eq_GeneralMapping} we have an $\NONLIN{\ADDn = n/7}{2^k}{\ADDd= \lceil 12/4 \rceil = 3}{2^3}$ additive code $\ADD$. The $\LIN{7}{3}{4}{2}$ Simplex code is a constant-weight code and therefore $\omega = 4$.
The code $\ADD$ is defined over $\F{2^3}$ and has dimension
\begin{align*}
\ADDk  =  k/3 = \frac{1}{3} \left( \frac{3}{7}n - m  \right) = \ADDn - \ADDext.
\end{align*}
and therefore the parameters of a $\LIN{((2^3)^{\ADDext}-1)/(2^3-1)}{\ADDn - \ADDext}{3}{2^3}$ Hamming code, which is optimal w.r.t. the sphere-packing bound.
\end{IEEEproof}
Construction~\ref{const_gclast} can be extended to the case where the locality code $\LOC$ is the $\LIN{2^a-1}{a}{2^{a-1}}{2}$ cyclic Simplex code, which is $2$-local and has availability $t= 2^{a-1}-1$ (see e.g. Kuijper--Napp~\cite[Lemma 3.1]{kuijper_erasure_2014} and Wang--Zhang~\cite{wang_repair_2014}).
\begin{construction}[Binary Code with Simplex Locality] \label{const_SimplexCode}
Let $n = 2^m-1$ and be divisible by $2^{a}-1$ and therefore $a|m$.
Let $\LOC$ be the $\LIN{2^a-1}{a}{2^{a-1}}{2}$ cyclic Simplex code with defining set 
\begin{equation}
\defset{\LOC} = \left\{ 0,\square,\square,3,\square,5,6,..,\square,2^{a-1}+1,..,2^a-1 \right\}.
\end{equation}
Let the defining set of the code $\CYC$ be:
\begin{align*}
\defset{\CYC} & = \left\{ \defset{\LOC} \cup \defset[2^a-1]{\LOC} \cup \defset[2(2^a-1)]{\LOC} \cup \dots \cup  \coset{1}{n} \right\}\\
 & = \big\{ \dots,-2^{a-1}+1,\dots,-1|,0,1,\dots,2^a-1,|2^a,\dots \big\}.
\end{align*}
Then $d \geq 2^a+2^{a-1}$ (via BCH bound for the consecutive sequence from $-(2^{a-1}-1)$ to $2^a$) and the dimension is $k =  \frac{a}{2^a-1}(2^m-1) - m$.
\end{construction}
We have the following theorem on the optimality of the dimension of linear codes.
\begin{theorem}[Simplex Locality] \label{theo_LinearSimplexCodeLocality}
Let $a|m$ and let $\CYC$ be an $\LIN{2^m-1}{k}{2^{a-1}\cdot 3}{2}$ linear code, let $\LOC$ be the $\LIN{2^a-1}{a}{2^{a-1}}{2}$ binary Simplex code and let $\CYC$ have the locality properties according to $\LOC$ as in Lemma~\ref{lem_ProjectionAdditiveCode}.
Then,
\begin{equation}
k \leq \frac{a}{2^a-1} \left( 2^m-1 \right)-m.
\end{equation}
\end{theorem}
\begin{IEEEproof}
From~\eqref{eq_GeneralMapping} we have an $\NONLIN{n/(2^a-1)}{2^k}{\lceil d/(2^{a-1}) \rceil }{2^a}$ additive code $\ADD$, where $\omega=2^a-1$, because the simplex code is a constant-weight code.
The additive code has the parameters of a Hamming code over $\F{2^a}$ with dimension
\begin{align*}
\ADDk & =  k/a = \frac{1}{a} \left( \frac{a}{2^a-1}n - m \right) = \ADDn - \ADDext,
\end{align*}
and distance
\begin{align*}
\ADDd  & =  \left \lceil \frac{d}{2^{a-1}} \right \rceil =  \left \lceil \frac{2^{a-1}(1+2)}{2^{a-1}} \right \rceil = 3.
\end{align*}
\end{IEEEproof}

\section{Q-Ary Case: First-Order Shortened RM Code as Locality Code} \label{sec_ReedMuller}
We extend the previous approach for cyclic codes to the $q$-ary case and use as locality code $\LOC$ the $q$-ary  
\begin{equation}
\LIN{q^2-1}{2}{(q-1)q^{2-1}}{q} = \LIN{q^2-1}{2}{q^2-q}{q}
\end{equation}
cyclic shortened first-order Reed--Muller (RM, see~\cite[Problem 2.17]{roth_introduction_2006} and \cite[Section 6.11]{van_lint_introduction_1999}) code. Its dual code is the $\LIN{q^2-1}{q^2-3}{2}{q}$ code with defining set $\{1,q\}$.
A $\LIN{q^a-1}{a}{q^{a-1}(q-1)}{q}$ shortened first-order RM code is the $q$-ary pendant of the Simplex code and also a constant-weight code with $\omega = q^{a-1}(q-1)$. RM codes have the highest minimum distance possible for the given parameters among $q$-ary linear codes. Furthermore, first-order RM codes and their locality properties were investigated by Rawat and Vishwanath in~\cite{rawat_locality_2012}.
\begin{construction}[Reed--Muller Code Locality] \label{const_QAryRM}
Assume $q > 2$. Let $\CYC$ be an $\LIN{q^m-1}{\CYCk}{\CYCd}{q}$ code.
Let $\LOC$ be an $\LIN{q^2-1}{2}{q^2-q}{q}$ cyclic RM code with defining set
\begin{equation*}
\defset{\LOC} = \left\{ 0,\square,2,3,\dots,q-1,\square,q+1,\dots,q^2-2 \right\}. 
\end{equation*}
Let the defining set of $\CYC$ be:
\begin{align*}
\defset{\CYC} & = \left\{ \defset{\LOC} \cup \defset[q-1]{\LOC} \dots \cup \{1,q,q^2,\dots,q^{m-1} \} \right\} \\
  & = \{ \square, (q^2-q-2),..,0,..,q^2+q-2,\square,..\}.
\end{align*}
Then the dimension of $\CYC$ is $k = \frac{2n}{q^2-1}-m$ and the distance is $d \geq q^2-q-2 + q^2+q-2 +1 +1 = 2q^2-2 $ via BCH bound~\cite{hocquenghem_codes_1959, bose_class_1960} (from $-(q^2-q-2) .. +(q^2+q-2)$).
\end{construction}
\begin{theorem}[Locality Code: Shortened First-Order RM Code] \label{theo_QAryConstructionRM}
Let $\CYC$ be an $\LIN{q^m-1}{k}{(q^2-q)\cdot 3}{2}$ linear code, let $\LOC$ be the $\LIN{q^2-1}{2}{q^2-q}{q}$ RM code and let $\CYC$ have the locality properties according to $\LOC$ as in Lemma~\ref{lem_ProjectionAdditiveCode}.
Then,
\begin{equation*}
k \leq \frac{2(q^m-1)}{q^2-1}-m.
\end{equation*}
\end{theorem}
\begin{IEEEproof}
The additive code $\ADD$ has parameters
\begin{align*}
\ADDn & = \frac{q^m-1}{q^2-1},\\
\ADDk & = k/2,\\
\ADDd & = \left \lceil \frac{d}{q^2-q} \right \rceil, 
\end{align*}
and has alphabet-size $\F{q^2}$.
More explicitly, the dimension is:
\begin{align*}
\ADDk = k/2 = \frac{1}{2} \left( \frac{2(q^m-1)}{q^2-1}-m \right) = \ADDn - \ADDext,
\end{align*}
and the distance is 
\begin{align*}
\ADDd & = \left \lceil \frac{d}{q^2-q} \right \rceil \geq \left \lceil \frac{2q^2-2}{q-1} \right \rceil  = \left \lceil \frac{q^2(2-2/q^2)}{q^2(1-\frac{1}{q})} \right \rceil = 3.
\end{align*}
Therefore the additive code has the parameters of an $q^2$-ary Hamming code, which is optimal w.r.t. to the sphere-packing bound.
\end{IEEEproof}
\begin{example}[Optimal Ternary Code] \label{ex_TernaryCodeRM}
Let $q = 3$ and let $n = 3^4-1 = 80$. Let $\LOC$ be the $\LIN{8}{2}{6}{3}$ shortened first-order cyclic RM code with defining set is $\defset{\LOC} = \{0,\square,2,\square,4,5,6,7\}$.
Then, the defining set of $\CYC$ according to Construction~\ref{const_QAryRM} is $\defset{\CYC}= \{ \{..,-4,..,0,\square,2,\square,4,..,8,\square,10,\square,..\} \cup \{1,3,9,27\} \}$ and therefore the BCH bound gives $d \geq 16$ ($-4..10$). And thus $\ADDd = \lceil 16/6 \rceil =3$. 
\end{example}
Construction~\ref{const_QAryRM} is also valid for extension fields and therefore let us give another example over a binary extension field.
\begin{example}[Cyclic Optimal Code over Binary Extension Field] \label{ex_BinaryExtension}
Let $q = 2^2$ and let $n = 4^4-1 = 255$.
Let $\LOC$ be the  $\LIN{15}{2}{12}{4}$ RM code with defining set $\defset{\LOC} = \{0,\square,2,3,\square,5,\dots,14\}$. 
Then, the defining set of $\CYC$ according to Construction~\ref{const_QAryRM} is $\defset{\CYC} = \{ \{..,-10,..,0,\square,2,3,\square,5,..,15,\square,17,18,\square,.. \} \cup \{1,4,16,64\} \}$.
The BCH bound gives $d \geq 30$ (from $-10..18$). The additive code over $\F{2^2}$ has length $\ADDn = 255/15 = 17$, dimension $\ADDk = 17-4 = 13$ and distance $\ADDd = \lceil 30/12  \rceil = 3$.
\end{example}
The real distance of the codes via Construction~\ref{const_QAryRM} is $3(q^2-q)$, but the BCH bound is not tight. Other bounds for cyclic codes can deliver a better result and a more advanced algebraic decoders for the non-local erasure-decoding can be applied.

\section{Optimal Linear Codes} \label{sec_LinearCodes}
Based on concatenated codes~\cite{forney_concatenated_1966, blokh_coding_1974}, we propose a construction, where the row-code is a linear Hamming code over the binary extension field $\F{2^r}$.
\begin{construction}[Binary Linear $r$-Local Code] \label{const_BinaryLinear}
Let the row code be a $\LIN{(2^{2r}-1)/(2^r-1) = 2^r+1}{2^r+1-2 = 2^r-1}{3}{2^r}$ binary Hamming code and let the column-code $\LOC$ be a linear (not necessarily cyclic) $\LIN{r+1}{r}{2}{2}$ single-parity check code.
Then the concatenated coded code is a $\LIN{(2^r+1)(r+1)}{(2^r-1)r}{6}{2}$ $r$-local code.
\end{construction}
Construction~\ref{const_BinaryLinear} gives an $r$-local linear code, with highest possible dimension $k$. (The additive codes is a binary Hamming code.)
\begin{example}[Binary Code with Locality $r=3$]
Let the $\LIN{((2^3)^2-1)/(2^3-1) = 9}{9-2 = 7}{3}{2^3}$ Hamming code be the row code of a concatenated code and let the column code be the $\LIN{4}{3}{2}{2}$ single-parity check code that corresponds to locality $r=3$.
Then $\CYC$ is a $\LIN{36}{21}{6}{2}$ $3$-local linear code.
The CM bound (Thm.~\ref{theo_CMBound}) gives $k \leq 21$.
\end{example}

\section{Conclusion and Outlook} \label{sec_conclusion}
We proposed new constructions of optimal binary and $q$-ary cyclic (and linear) codes with locality and availability.

The following future work seems fruitful. The extension of Construction~\ref{const_SimplexCode} to codes with higher distance and optimal dimension, the usage of other first-order cyclic Reed--Muller codes as locality code similar to Construction~\ref{const_QAryRM}, the usage of improved bounds on the minimum distance for the cyclic codes obtained via Construction~\ref{const_QAryRM}  and the extension of Construction~\ref{const_BinaryLinear} to $q$-ary linear codes.

\end{document}